\documentclass[12pt]{iopart}
\usepackage{graphicx}
\usepackage{bm}

\begin{document}

\title[Statistical and geometrical properties of thermal plumes in turbulent RBC]{Statistical and geometrical properties of thermal plumes in turbulent Rayleigh-B\'{e}nard convection}

\author{Quan Zhou$^{1,2}$ and Ke-Qing Xia$^1$}

\address{$^1$Department of Physics, The Chinese University of Hong Kong, Shatin, Hong Kong, China \\
$^2$Shanghai Institute of Applied Mathematics and Mechanics, Shanghai University, Shanghai 200072, China}
\ead{kxia@phy.cuhk.edu.hk}

\begin{abstract}
We present a systematic experimental study of geometric and
statistical properties of thermal plumes in turbulent
Rayleigh-B\'{e}nard convection using the
thermochromic-liquid-crystal (TLC) technique. The experiments were
performed in three water-filled cylindrical convection cells with
aspect ratios 2, 1, and 0.5 and over the Rayleigh-number range
$5\times10^7 \leq Ra \leq 10^{11}$. TLC thermal images of
horizontal plane cuts at various depths below the top plate were
acquired. Three-dimensional images of thermal plumes were then
reconstructed from the two-dimensional slices of the temperature
field. The results show that the often-called sheetlike plumes are
really one-dimensional structures and may be called rodlike
plumes. We find that the number densities for both
sheetlike/rodlike and mushroomlike plumes have power-law
dependence on $Ra$ with scaling exponents of $\sim 0.3$, which is
close to that between the Nusselt number $Nu$ and $Ra$. This result
suggests that it is the plume number that primarily determines the
scaling exponent of the $Nu$-$Ra$ scaling relation. The evolution
of the aspect ratio of sheetlike/rodlike plumes reveals that as
$Ra$ increases the plume geometry changes from more-elongated to
less-elongated. Our study of the plume area fraction (fraction of
coverage over the surface of the plate) further reveals that the
increased plume numbers with $Ra$ mainly comes from increased
plume emission, rather than fragmentation of plumes. In addition,
the area, perimeter, and the shape complexity of the
two-dimensional horizontal cuts of sheetlike/rodlike plumes were studied and
all are found to obey log-normal distributions.
\end{abstract}

\maketitle

\section{Introduction}

As an important class of turbulent flows, turbulent thermal
convection is ubiquitous in nature, ranging from those in the
planets and stars, in the Earth's mantle and its outer core, and
in the atmosphere and oceans to the convection in heat transport
and mixing in engineering applications. Turbulent
Rayleigh-B\'{e}nard (RB) convection, a fluid layer sandwiched
between two parallel plates and heated from below, has long been
used as a model system to study natural convections
\cite{siggia1994arfm, agl, lx}. Two of the important issues in the
study of turbulent RB convection are heat transport and coherent
structures such as thermal plumes. In the first one, one tries to
understand how heat transported upwards across the fluid layer,
which is characterized by the Nusselt number
$Nu=J/(\chi\bigtriangleup/H)$, depends on the turbulent
intensity, which is characterized by the Rayleigh number
$Ra=\alpha g\Delta H^{3}/\nu\kappa$. Here, $\Delta$ is the applied
temperature difference across the fluid layer, $g$ is the
gravitational acceleration, and $\alpha$, $\nu$, $\chi$ and
$\kappa$ are, respectively, the volume expansion coefficient,
kinematic viscosity, thermal conductivity, and thermal diffusivity
of the convecting fluid. In many theories and experiments, it is
often assumed that there is a simple power law relation between
$Nu$ and $Ra$, i.e. $Nu\sim Ra^{\beta}$. In the second issue one
wants to understand the statistical and geometrical properties of
thermal plumes, which is a localized thermal structure and has
been shown to play a key role in many natural phenomena and
engineering applications, such as in mantle convection where
mantle plumes just below Earth's crust are responsible for the
formation of volcanoes (see, for example, Ref.
\cite{morgan1968jgr, jellinek2004rog}), in nuclear explosions and
stellar convection where plumes dominate both the dynamics and the
energy transport (see, for example, Ref. \cite{rieutord1995aaa}).
The two issues are not independent of each other. It has been
shown recently that for turbulent RB system most heat is carried
and transported by thermal plumes \cite{shang2003prl,
shang2004pre, pinton2007prl}. Accordingly, it remains a challenge
to establish a quantitative relationship between thermal plumes
and heat transport. Although a theoretical effort \cite{gl2004pof}
has recently focused on this connection, few such experimental
studies have been made.

Thermal plumes generated by a localized hot/cold spot have a well
organized structure, which consists of a mushroom cap with sharp
temperature gradient and a stem that is relatively diffusive
\cite{moses1993jfm, kaminski2003jfm}. Plumes with such a structure
are often referred to as mushroomlike plumes, which are also
observed in turbulent RBC when viewed from the side
\cite{zocchi1990pa, moses1991epl, gluckman1993pof, moses1993jfm,
ciliberto1996pre, zhang1997pof, du1998prl, qiu2001prl,
shang2003prl, xi2004jfm}. By extracting plumes from temperature
time series measured locally near the cell sidewall, Zhou and Xia
\cite{zhou2002prl, zhou2009} showed that the size of mushroomlike
plumes obey log-normal statistics. However, the morphology of the
plumes is totally different when one observes from above (or
below): thermal plume is extended in one horizontal direction but
concentrated in the orthogonal horizontal direction
\cite{tanaka1980ijhmt, zocchi1990pa, gluckman1993pof,
vorobie2002jfm, funfschilling2004prl, haramina2004pre,
puthenveettil2005jfm1, puthenveettil2005jfm2, zhou2007prl,
funfschilling2008jfm, puthenveettil2008jfm}. Plumes with such a
structure are often assumed to have significant vertical extent
and thus are called sheetlike plumes. Puthenveettil and Arakeri
\cite{puthenveettil2005jfm1} studied near-wall structures in
turbulent natural convection driven by concentration differences
across a membrane and found that the plume spacings show a common
log-normal probability density function (PDF). Zhou \emph{et al.}
\cite{zhou2007prl} further revealed that both the area and the
heat content of sheetlike plumes are log-normal distributed.
Shishkina and Wagner \cite{shishkina2008jfm} investigated
quantitatively geometric properties of sheetlike plumes using
direct numerical simulations. However, systematical experimental
studies of geometrical structures of sheetlike plumes are still
missing. Recently, Funfschilling \emph{et al.}
\cite{funfschilling2008jfm} suggested that when viewed from above
thermal plumes near the two plates should be better referred to as
linelike plumes, as the vertical extent of these structures do not
seem to be established and it appears more likely that they are
one dimensional excitations in the marginally-stable boundary
layers.

As two different configurations of thermal plumes coexist
simultaneously in turbulent RB system, it is natural to ask how
sheetlike and mushroomlike plumes transform from each other. By
using thermochromic-liquid-crystal (TLC) technique, Zhou \emph{et
al.} \cite{zhou2007prl} showed, in a cylindrical cell with unity
aspect ratio ($\Gamma=D/H$ with $D$ and $H$ as the inner diameter
and the height of the convection cell, respectively.), that hot
fluids (plumes) move upwards, imping on the top plate from
below, then spread horizontally along the top plate and form waves
or sheetlike plumes. As they travel horizontally along the plate,
these sheetlike plumes collide with each other or with the
sidewall, convolute and form swirls. As these swirls are cooler
than the bulk fluid, they move downwards, merge and cluster
together. By the symmetry of the system, the same process is
expected for the morphological evolution of thermal plumes
occurring near the bottom plate. However, in Ref.
\cite{zhou2007prl} the morphological evolution was visualized only
at a single Rayleigh number ($Ra=2.0\times10^9$), and it is not
clear whether this process is universal for much higher $Ra$ or
for cells with different aspect ratios. The understanding of this
evolution process is of great importance, as the characterization
of coherent structures is essential to the understanding of
turbulent flows in many systems.

In the present paper, we report new experiments of the temperature
and velocity fields measured at varying depths from the top plate
over the Rayleigh number range $5\times10^7 \leq Ra \leq 10^{11}$
and the Prandtl number range $4.1 \leq Pr = \nu/\kappa \leq 5.3$ and in three
water-filled cylindrical sapphire cells with aspect ratios 0.5, 1,
and 2. We present quantitative results on the relationship between
plume number density and $Nu$ and results that relate the
evolution of the plume morphology to the heat transport. The
remainder of the paper is organized as follows. We give a detailed
description about the experimental setup and data analysis method
in section 2. In section 3, we study the $z$- and $Ra$-dependence
of plume number and discuss the relationship between thermal
plumes and the $Nu$-$Ra$ scaling. Section 4 presents study on the
geometric properties of sheetlike plumes, which are mainly based
on data from the aspect ratio 1 cell. We summarize our findings
and conclude in section 5.

\section{Experimental setup and procedures}

\subsection{The convection cell and the experimental parameters}

\begin{figure}
\begin{center}
\includegraphics[scale=0.7]{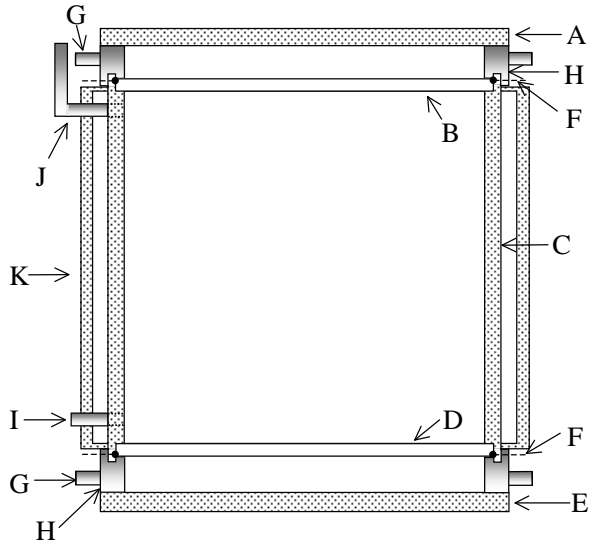}
\end{center}
\caption {A schematic drawing of the convection cell used in the
experiments. The top  and bottom plates are made of sapphire so
that the flow in a horizontal plane can be visualized and
captured. A: the cooling chamber cover, B: the top sapphire plate,
C: the Plexiglas sidewall, D: the bottom sapphire plate, E: the
heating chamber cover, F: thermistors, G: nozzles for cooling or
heating water, H: stainless steel rings as part of the heating and
cooling chambers, I: nozzle for filling fluid into the cell, J:
nozzle for letting air out of the cell, K: a square-shaped glass
jacket.}\label{fig1}
\end{figure}

The experiments were carried out in three cylindrical sapphire
cells \cite{xi2006pre, zhou2007prl}. The schematic diagram of the
cells is shown in figure \ref{fig1}. To capture and study the
horizontal temperature and velocity fields from top, two sapphire
discs (Almaz Optics, Inc) with diameter $19.5$ cm and thickness
$5.0$ mm were chosen as the top and bottom plates for their good
thermal conductivity ($35.1$ W/mK at 300 K) compared with other
transparent materials. Two chambers, constructed by stainless
steel rings $H$ and Plexiglas discs $A$ and $E$, are used to heat
the bottom plate and cool the top plate. Each chamber is connected
to a separate refrigerated circulator by four nozzles, water flows
into the chamber by two nozzles in two opposite direction and
leaves from the other two nozzles perpendicular to the inlets. The
sidewall of the cell is a vertical tube made of Plexiglas ($C$)
with inner diameter $D=18.5$ cm and wall thickness $8$ mm,
respectively. The separations between the top and bottom plates
$H$ are $9.3$, $18.5$, and $37$ cm so that the aspect ratios of
the cells are respectively $\Gamma=2$, 1, and 1/2. Draining ($I$)
and filling ($J$) tubes are fitted on the Plexiglas tube at a
distance of $1.5$ cm from the top and bottom plates.  A
square-shaped jacket ($K$) made of flat glass plates and filled
with water is fitted to the outside of the sidewall
\cite{xi2004jfm}, which greatly reduces the spatial variation in
the intensity of the white lightsheet caused by the curvature of
the cylindrical sidewall. Four rubber O-rings (not shown in the
figure) are placed between the two sapphire plates and the steel
rings and between the rings and the two Plexiglas discs to avoid
fluid leakage. Four thermistors $F$ (model 44031, Omega
Engineering Inc.) are used to measure the temperature difference
between the two sapphire plates. To keep good contact between the
plates and the thermistors, the thermistors are wrapped by the
heat transfer compound (HTC10s from Electrolube Limited). It is
found that the measured relative temperature difference between
the two thermistors in the same plate is less than $3\%$ of that
across the convection cell for both plates and for all $Ra$
investigated. This indicates the uniform distribution of the
temperature across the horizontal plates.

\subsection{Liquid crystal measurements}

\begin{figure}
\begin{center}
\includegraphics[scale=1]{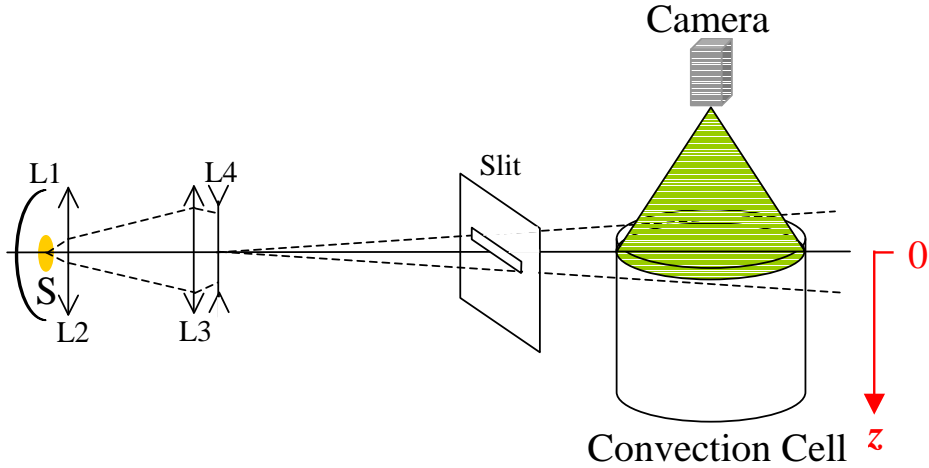}
\end{center}
\caption {The optical setup for thermochromatic liquid crystal
visualization and measurements: $S$, halogen lamp; $L1$, concave
mirror; $L2$ and $L3$, condensing lenses; $L4$, diverging
cylindrical lens.}\label{fig2}
\end{figure}

The visualization technique employing thermochromatic liquid
crystal (TLC) particles has been widely used and documented in
fluid visualization experiments and was used in the present work
to visualize the temperature and velocity fields in horizontal
fluid layers of varying depth from the cell's top plate. Two types
of TLC microspheres (Hallcrest, Ltd.) were used in the
experiments. One type (model R29C4W) was used for low-$\Delta$
experiments with the mean bulk temperature $T_0=30$ $^{\circ}$C
and $Pr=5.3$, and the other type (model R40C5W) was used for
high-$\Delta$ measurements with $T_0=42$ $^{\circ}$C and $Pr=4.1$.
Both these particles have a mean diameter of $50$ $\mu$m and
density of $1.03-1.05$ g/cm$^3$, and were suspended in the
convection fluid in very low concentrations (about $0.01\%$ by
weight), at which the influence of TLC particles on the fluid can
be neglected. The peak wavelength of light scattered by these
particles changes from red to green and then to blue within a
temperature window of 4 $^{\circ}$C from about 29 to 33
$^{\circ}$C for R29C4W, and of 5 $^{\circ}$C from about 40 to 45
$^{\circ}$C for R40C5W. In the experiments, the mean bulk
temperature was set to 30 $^{\circ}$C for R29C4W (Pr $= 5.3$) and
to 42 $^{\circ}$C for R40C5W (Pr $= 4.1$), so that the background
fluid appears blue and the red and green regions correspond to
cold fluid, i.e. cold plumes. Figure \ref{fig2} shows a schematic
diagram of the optical setup for the experiments. A halogen photo
optic lamp ($S$) with a power of 650 W was used as the light
source. One concave mirror $L1$ and two condensing lenses $L2$ and
$L3$ were used to collect the light from $S$ and focus it onto the
central section of the cell. A horizontal sheet of white light,
generated by a diverging cylindrical lens $L4$ and then projected
onto an adjustable slit, passed through the cell parallel to the
top plate. The thickness of the lightsheet inside the cell is
approximately 3 mm. A Nikon D1X camera, with a resolution
$2000\times1312$ pixels and 24 bit dynamic range, was placed on
the top of the cell to take photographs of the TLC microspheres.
With short camera exposure time ($0.02$ s), the captured
photographs give the instantaneous temperature field, and with
long camera exposure time (0.77s) they will in addition show the
the trajectories of the particles.

\begin{figure}
\begin{center}
\includegraphics[scale=1]{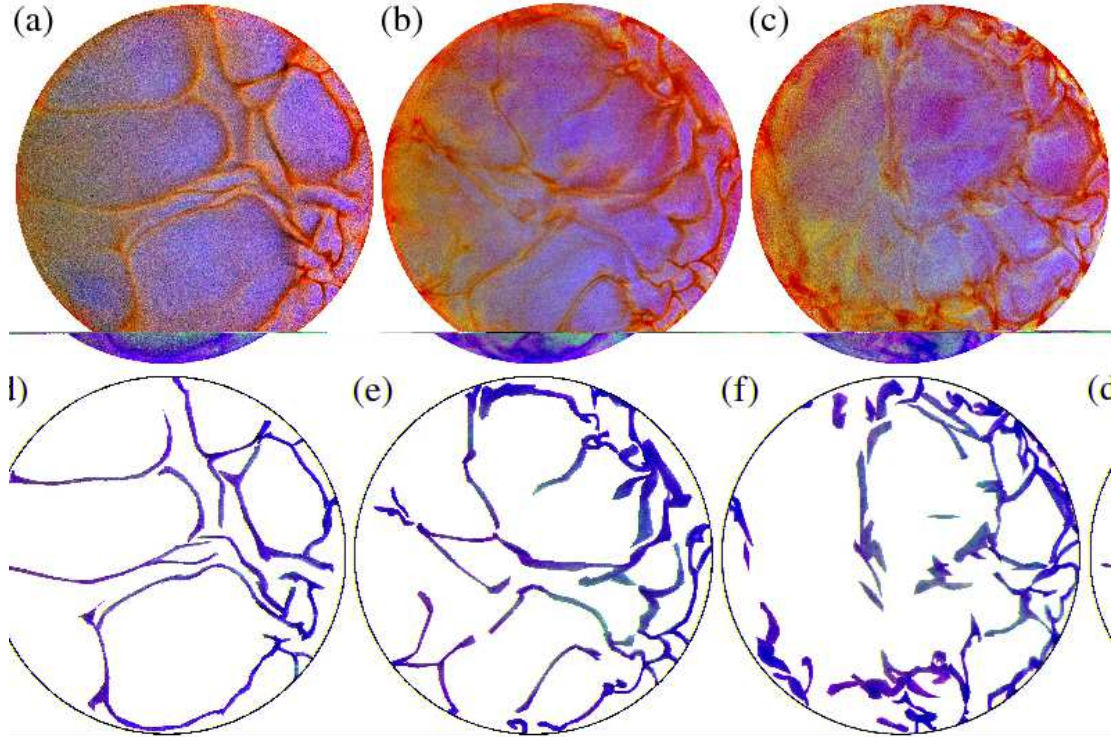}
\end{center}
\caption {Images of TLC microspheres with the camera exposure time
of 0.02 s taken  at $2$ mm from the top plate at (a)
$Ra=1.1\times10^9$, (b) $3.0\times10^9$, and (c)
$1.0\times10^{10}$. (d), (e) and (f), which correspond to (a), (b)
and (c), respectively, are extracted sheetlike plumes with
background in the original images removed.}\label{fig3}
\end{figure}

\begin{figure}
\begin{center}
\includegraphics[scale=0.85]{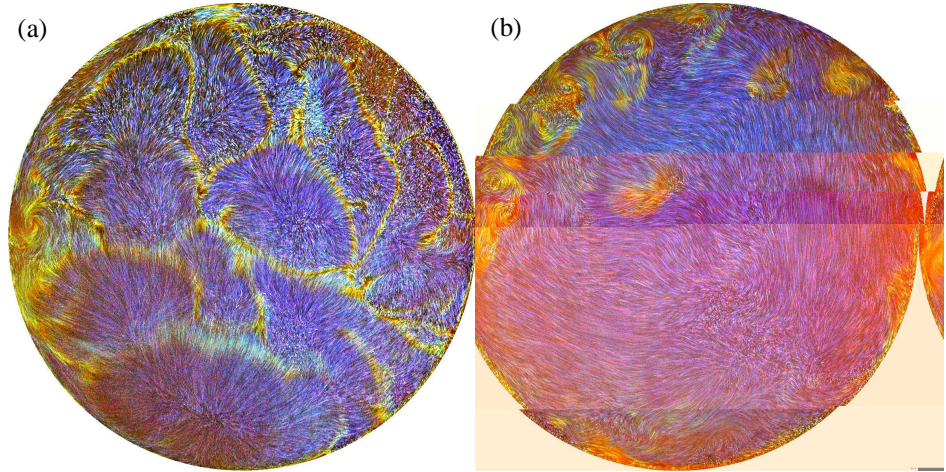}
\end{center}
\caption {Images of TLC microspheres with the camera exposure time
of 0.77 s taken at (a) $2$ mm and (b) $2$ cm from the top plate at
$Ra=1.0\times10^9$.}\label{fig4}
\end{figure}

To study thermal plume properties quantitatively, we counted plume
numbers and extracted sheetlike plumes. For sheetlike plumes, we
took $150$ to $300$ consecutive images (at $z=2$ mm and with
camera exposure time of $0.02$ s) at $30$ to $60$ s intervals for
a given $Ra$, so that two successive images in each sequence are
statistically independent. A cold sheetlike plume is extracted by
first manually drawing a contour around its perimeter and then
using a software to collect all pixels enclosed by the contour. A
program is then used to calculate the perimeter $P$ and the area
$A$ of each extracted plume. To ensure that plumes are identified
correctly the operator uses knowledge gained from viewing movies
of plume motions. A total of $3000$ to $6000$ plumes are
identified from the $150$ to $300$ images, which are not large
numbers but should be indicative of the statistical properties.
Figures \ref{fig3}(d)-(f) show three examples of extracted plumes
with background removed, which correspond to images of TLC
microsperes in figures \ref{fig3}(a)-(c), respectively, for three
different values of $Ra$ obtained from the unity-aspect-ratio
cell. For mushroomlike plumes, we took sequences of images at
varying depth $z$ from the top plate and with camera exposure time
of $0.77$ s at $30$ s intervals for each $Ra$. Each sequence
consists $200$ images for a given depth. A cold mushroomlike plume
was identified as an object with nonzero vorticity and with
temperature (color) much lower than that of the background fluid
\cite{zhou2007prl} (see, e.g., figure \ref{fig4}(b)). As
mushroomlike plumes are partly entangled together, they cannot be
separated completely. Nevertheless, we can identify them
individually as swirls with cold temperature and thus count their
numbers. The cold sheetlike plumes were also identified and
counted from these images, as objects with a linelike shape and
lower temperature.

\section{Plume number statistics}

\subsection{Morphological evolution of thermal plumes}

\begin{figure}
\begin{center}
\includegraphics[scale=0.75]{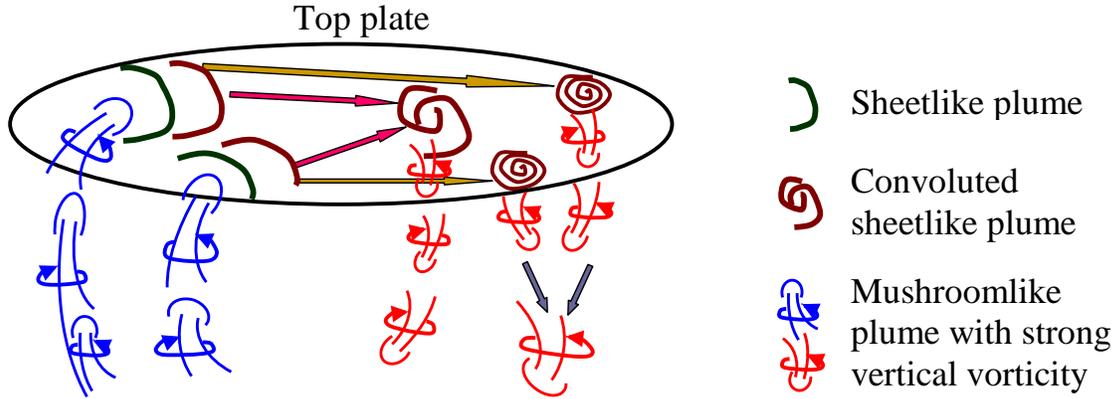}
\end{center}
\caption {The process of morphological evolution between sheetlike
and mushroom plumes.}\label{fig5}
\end{figure}

The process of morphological evolution between sheetlike and
mushroomlike plumes has been revealed by Zhou \emph{et al.}
\cite{zhou2007prl} in a cylindrical cell. Here, the same process
was also observed for all values of $Ra$ and $\Gamma$ investigated
in the experiment. Figure \ref{fig4}(a), taken at 2 mm from the
top plate, shows how this evolution comes about. One sees that
near the top plate the motion of TLC microspheres appear to
emanate from certain regions or ``sources" with bluish color,
suggesting that hot fluids (plumes) are moving upwards, impinging
on and spreading horizontally along the top plate. Along the
particle traces, the color turns from blue to green and red,
implying that the wave fronts are cooled down gradually by the top
plate (and the top thermal boundary layer) as they spread. As they
travel along the plate's surface, sheetlike plumes collide with
each other or with the sidewall. As different plumes carry momenta
in different directions, they merge, convolute and form swirls
(hence generating vorticity). As these swirls are cooler than the
bulk fluid, they move downwards, merge and cluster together (see,
e.g, figure \ref{fig4}(b)). We note that, by symmetry of the
system, the same morphological evolution of thermal plumes should
occur near the bottom plate. The physical picture of this
evolution process is illustrated in figure \ref{fig5}.

\subsection{Depth-dependent properties}
\label{sec:z}

\begin{figure}
\begin{center}
\includegraphics[scale=1]{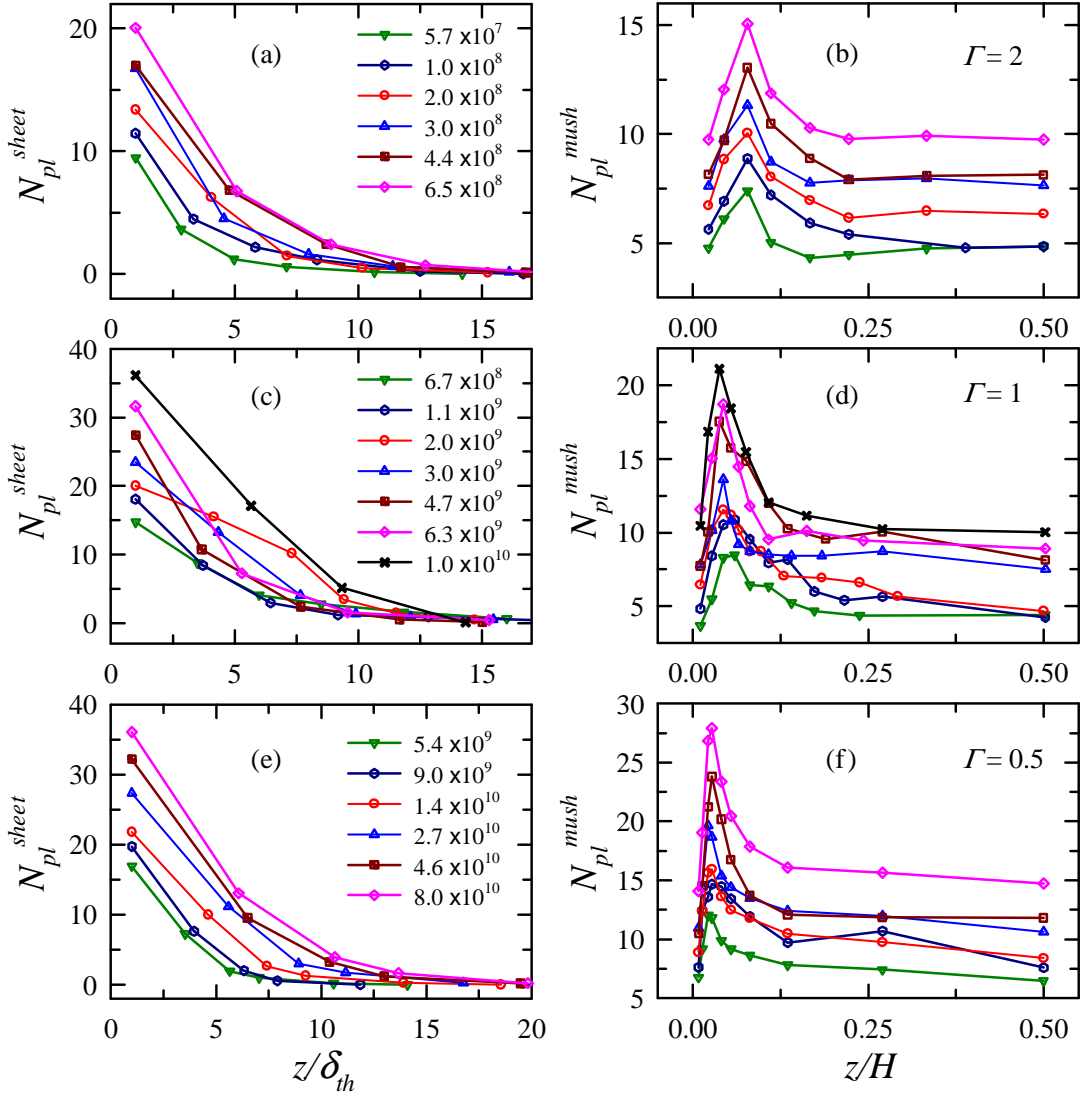}
\end{center}
\caption {Mean numbers of (a), (c), and (e) (cold) sheetlike
plumes $N_{pl}^{sheet}$  and (b), (d), and (f) (cold) mushroomlike
plumes $N_{pl}^{mush}$ as functions of the distance $z$ for all
measured $Ra$. (a) and (b) for $\Gamma=2$, (c) and (d) for
$\Gamma=1$, and (e) and (f) for $\Gamma=0.5$.}\label{fig6}
\end{figure}

\begin{figure}
\begin{center}
\includegraphics[scale=1]{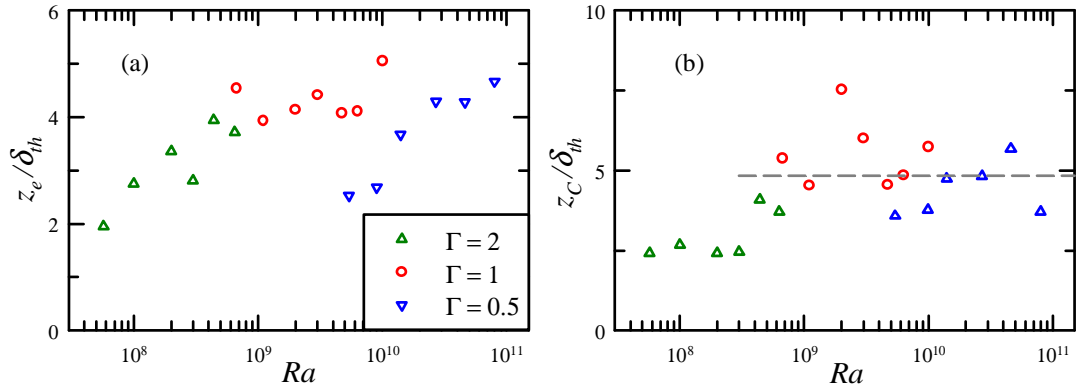}
\end{center}
\caption {(a) $z_e/\delta_{th}$ and (b) $z_C/\delta_{th}$ vs $Ra$
for $\Gamma=2$  (up-triangles), 1 (circles), and 0.5
(down-triangles). The dashed line in (b) marks the mean value of
$z_c$ for $Ra>3\times10^8$.}\label{fig7}
\end{figure}

Because of the collision and convolution of sheetlike plumes and
the transformation from sheetlike to mushroomlike ones, it is
expected that the number of sheetlike plumes should decrease from
the top plate, while that of mushroomlike ones should increase,
and this is indeed the case as shown in figure \ref{fig6}. Figures
\ref{fig6}(a), (c), and (e) show the mean numbers of (cold)
sheetlike plumes $N_{pl}^{sheet}$ versus the normalized height
$z/\delta_{th}$ obtained from the cells with $\Gamma=2$, 1, and
0.5, respectively. (Here, $\delta_{th}$ is the thickness of
thermal boundary layer, which was measured in \cite{lui1997,
lui1998pre}.) One sees that $N_{pl}^{sheet}$ decays quickly away
from the top plate and the behavior is similar for all $Ra$
studied. Beyond the depth $z\approx15\delta_{th}$, one can hardly
identify any sheetlike plumes from the acquired images. Note that,
in addition to transforming into mushroomlike plumes, the rapid
reduction of $N_{pl}^{sheet}$ with depth could also be attributed
to the limited vertical extent of sheetlike plumes themselves,
i.e. plumes, whose vertical extents are shorter than a certain
height $h$, could not be captured by the images that are taken
farther than $h$ from the top plate. We further note that the
$z$-dependence of plume numbers $N_{pl}^{sheet}$ may be well
described by a decreasing exponential function, i.e.
$N_{pl}^{sheet}=N_0 e^{-b_r z/\delta_{th}}$ with fitting
parameters $N_0$ and $b_r$ for nearly all measured $Ra$. Hence
$z_e=1/b_r$ may be used as a typical height of vertical extent of
sheetlike plumes. Figure \ref{fig7}(a) shows the scaled height
$z_e/\delta_{th}$ as a function of $Ra$. It is seen that
$z_e/\delta_{th}$ increases with $Ra$ for the $\Gamma=2$ and 0.5
cells, while keeps nearly constant for the $\Gamma=1$ cell. We can
not judge the significance of the different behaviors of
$z_e/\delta_{th}$ varying with $Ra$ in cells with different
$\Gamma$. However, $z_e$ varies only in the range of a few times
the thickness of thermal boundary layer
($2\delta_{th}<z_e<5\delta_{th}$). This suggests that the vertical
extent of the so-called sheetlike plumes does not have enough
spatial extension in the vertical direction to form sheets.

For mushroomlike plumes, the mean numbers of (cold) mushroomlike
plumes $N_{pl}^{mush}$ as a function of the scaled depth $z/H$ is
shown in figures \ref{fig6}(b), (d), and (f). With increasing $z$,
$N_{pl}^{mush}$ for all $Ra$ first increases rapidly, then
decreases and finally remains approximately constant for positions
$z>0.2H$. When $N_{pl}^{mush}$ first increases with $z$, it
crosses over with $N_{pl}^{sheet}$ at some crossover-depth $z_C$.
The normalized crossover depths $z_C$ are shown in figure
\ref{fig7}(b). It is seen that $z_c\approx2.5\delta_{th}$ when
$Ra\leq3\times10^8$. For $Ra>3\times10^8$, although the data
points seem to be somewhat scattered, they exhibit no clear
dependence on $Ra$ and have a mean value of $\sim4.84\delta_{th}$.
This implies that the transformation occurs within a region, whose
height is only associated with the thermal boundary layer
thickness. If one takes this value as the typical vertical extent
of sheetlike plumes, these results again suggest that the vertical
extent of what is called sheetlike plumes is only the order of
several $\delta_{th}$, not much larger than it, and thus not large
enough to have a sheetlike shape. This will be further discussed
in section $\ref{sec:3D}$. After $N_{pl}^{mush}$ attains its
maximum value around $z=z_p$, it then drops sharply beyond the
depth $z\simeq0.2H$, because of the mixing, merging, and
clustering of mushroomlike plumes. As found by Zhou \emph{et al.}
\cite{zhou2007prl}, this region corresponds to the region of full
width at half maximum of vertical vorticity profile and can be
used as a quantitative definition and measure of the mixing zone
\cite{castaing1989jfm, zhou2002prl}. It is further found that the
peak position in the profiles of $N_{pl}^{mush}$, $z_p$, has no
obvious $Ra$-dependence for each $\Gamma$, but decreases slightly
from $0.08H$ for $\Gamma=2$ to $0.03H$ for $\Gamma=0.5$.


\subsection{The Rayleigh number dependency}


\begin{table}
\caption{The fitted parameters $\alpha_s$, $\beta_s$,
$\alpha_m$, and $\beta_m$ for
all three cells.}\label{tab1} \footnotesize\rm
\begin{tabular*}{\textwidth}{@{}l*{15}{@{\extracolsep{0pt plus12pt}}l}}
\br
$\Gamma$ & $\alpha_s$ & $\beta_s$ & $\alpha_m$ & $\beta_m$\\
\mr
0.5 & 1.2 & $0.29\pm0.03$ & 0.45 & $0.31\pm0.03$ \\
1 & 1.5 & $0.28\pm0.03$ & 0.37 & $0.33\pm0.03$ \\
2 & 1.6 & $0.30\pm0.03$ & 1.86 & $0.28\pm0.03$ \\
\br
\end{tabular*}
\end{table}

\begin{figure}
\begin{center}
\includegraphics[scale=0.7]{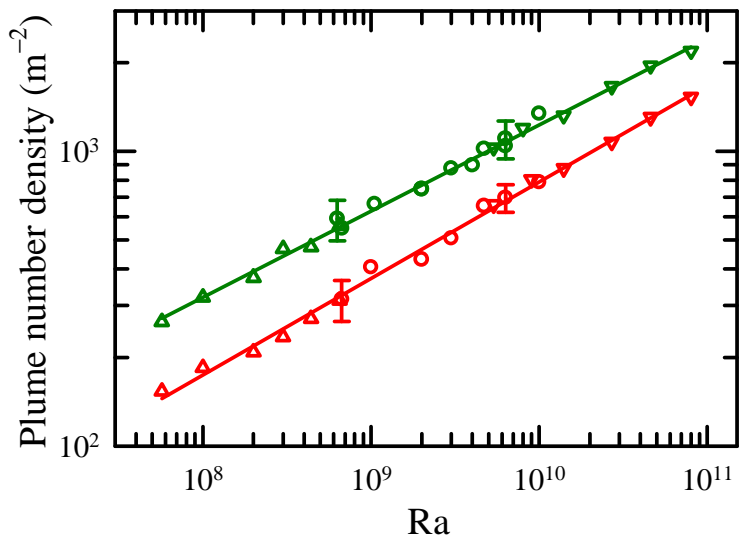}
\end{center}
\caption {$Ra$-dependency of sheetlike plume number density
$N_{pl}^{sheet}/(\pi D^2/4)$ (dark-green symbols), counted just
outside the thermal boundary layer, and mushroomlike plume numbers
$N_{pl}^{mush}/(\pi D^2/4)$ (red symbols), counted at the peak of
the height-profile of $N_{pl}^{mush}$, for the $\Gamma=2$
(up-triangles), 1 (circles), and 0.5 (down-triangles) cells. For
both $N_{pl}^{sheet}$ and $N_{pl}^{mush}$, the $\Gamma=2$ (0.5)
data have been shifted down (up), so that they agree with
corresponding data for $\Gamma=1$ in their overlap regions of
$Ra$. Solids lines are power-law fittings to the data shown in the
same color. The fitted scaling exponents are $0.29\pm0.03$ and
$0.32\pm0.03$ for $N_{pl}^{sheet}$ and $N_{pl}^{mush}$,
respectively.}\label{fig8}
\end{figure}

We next focus on the $Ra$-dependency of plume number density. For
sheetlike plumes, the number was counted just outside the thermal
boundary layer and the counted number, $N_{pl}^{sheet}$, for each
cell can be well described by a power law $N_{pl}^{sheet}/(\pi
D^2/4)=\alpha_s(\Gamma) Ra^{\beta_s(\Gamma)}$, with two fitting
parameters $\alpha_s$ and $\beta_s$ as functions of $\Gamma$.
Here, the fitted values of $\alpha_s$ and $\beta_s$ for all three
aspect ratios are listed in table \ref{tab1}. It is seen that
$\alpha_s(\Gamma)$ varies with the cell's aspect ratio $\Gamma$
while $\beta_s(\Gamma)$ is approximately the same for all three
cells. As the range of $Ra$ is limited for each aspect ratio, we
shifted the plume number densities for the $\Gamma=2$ and 0.5
cells to agree with data for the $\Gamma=1$ cell in their
respective overlapping range of $Ra$. We then fitted a single
power-law to the combined data set from the three cells that span
almost three decades of $Ra$. Figure \ref{fig8} shows the shifted
numbers, normalized by the area of the top plate $\pi D^2/4$, as
well as the plume number from the $\Gamma=1$ cell as dark-green
symbols. It is seen that a single power law
\begin{equation}
\label{eq:nsheet}
N_{pl}^{sheet}/(\pi D^2/4)=\alpha_s Ra^{\beta_s} \mbox{\ \ with\ \ } \alpha_s=1.4 \mbox{\ \ and\ \ } \beta_s=0.29\pm0.03
\end{equation}
can be used to well describe the data from all three cells.


For mushroomlike plumes, one sees from figures \ref{fig6}(b), (d),
and (f) that there is a peak on the  height-profile of
mushroomlike plume number $N_{pl}^{mush}$ for all three cells and
for all measured $Ra$. We use the value at the peak of the
height-profile of mushroomlike plume number to examine the
$Ra$-dependence of $N_{pl}^{mush}$. It is found that a power law
relation $N_{pl}^{mush}/(\pi D^2/4)=\alpha_{m}(\Gamma)
Ra^{\beta_{m}(\Gamma)}$ can also be used to describe the
$Ra$-dependence of $N_{pl}^{mush}$. Again, the fitted
$\beta_{m}(\Gamma)$ is nearly $\Gamma$-independent (see table
\ref{tab1}). Figure \ref{fig8} shows the $Ra$-dependence of
$N_{pl}^{mush}/(\pi D^2/4)$ (red symbols). Here, the $\Gamma=2$
and 0.5 data have also been shifted to agree with the $\Gamma=1$
in their respective overlap regions of $Ra$. It is seen that
$N_{pl}^{mush}$ shows similar trend with $Ra$ as $N_{pl}^{sheet}$.
Again, the relationship between $N_{pl}^{mush}$ and $Ra$ can be
well represented by power-law fits
\begin{equation}
\label{eq:nmp}
N_{pl}^{mush}/(\pi D^2/4) = \alpha_{m}Ra^{\beta_{m}} \mbox{\ \ with\ \ } \alpha_{m}=0.41 \mbox{\ \ and\ \ } \beta_{m}=0.32\pm0.03.
\end{equation}

To make a quantitative connection between thermal plumes and the
heat flux, we use the previous experimental finding from both
Eularian and Lagrangian measurements that the heat flux are mainly
carried by thermal plumes \cite{shang2003prl, pinton2007prl}. As
mushroomlike plumes are evolved morphologically from sheetlike
plumes \cite{zhou2007prl}, the total heat flux carried by either
type should be the same, i.e.,
\begin{equation}
Nu\simeq N_{pl}^{sheet} F_{pl}^{sheet} \simeq N_{pl}^{mush} F_{pl}^{mush} \sim Ra^{\beta}.
\end{equation}
Here, $F_{pl}^{sheet}$ and $F_{pl}^{mush}$ are the mean heat flux
carried by individual sheetlike and mushroomlike plumes,
respectively, and $\beta\simeq0.3\pm0.02$ in the $Ra$ and $Pr$
ranges of the experiment \cite{xia2002prl, ahlers2003prl}. Recall
that the $Ra$-dependence of $N_{pl}^{sheet}$ [Eq.
(\ref{eq:nsheet})] and $(N_{pl}^{mush})_{peak}$ [Eq.
(\ref{eq:nmp})] discussed above, one sees that the scaling
exponents of the $Nu$-$Ra$ relation and of the $Ra$-dependence of
plume numbers are the same within experimental uncertainty. This
indicates that the heat flux transported by individual plumes is
nearly independent of the turbulent intensity, i.e. $Ra$, and
hence the $Nu$-$Ra$ scaling relation is determined primarily by
the number of thermal plumes.

\section{Geometric properties of sheetlike plumes}

\subsection{Aspect ratio of sheetlike plumes}

\begin{figure}
\begin{center}
\includegraphics[scale=1]{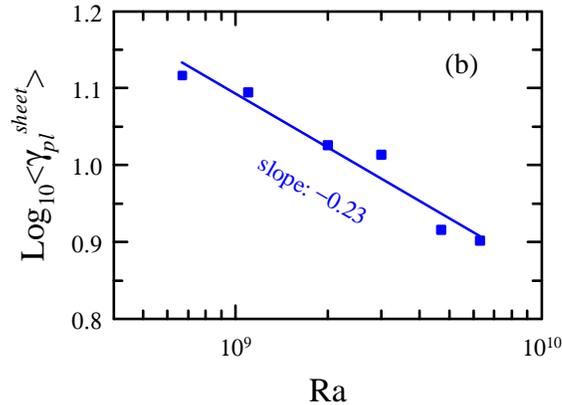}
\end{center}
\caption {$Ra$-dependence of the mean aspect ratio of sheetlike
plumes $\langle\gamma_{pl}^{sheet}\rangle$ vs. $Ra$ in a log-log plot.} \label{fig9}
\end{figure}

\begin{figure}
\begin{center}
\includegraphics[scale=1]{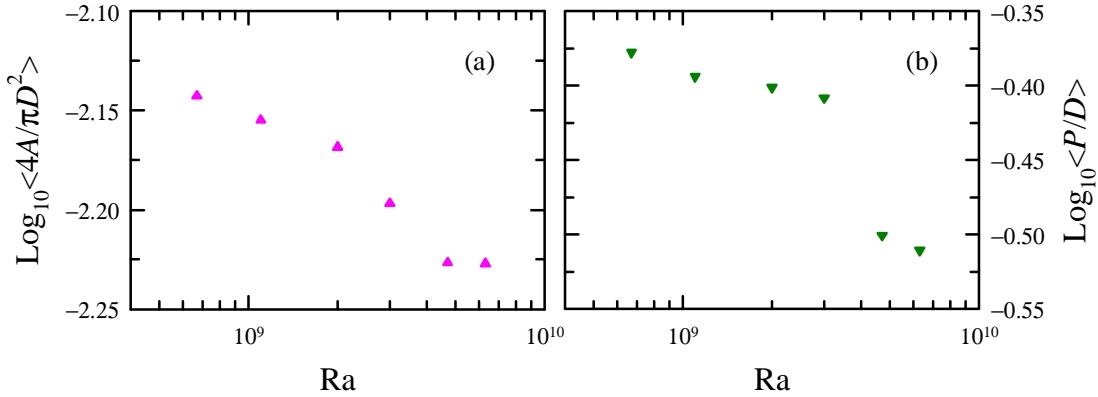}
\end{center}
\caption {$Ra$-dependence of (a) the mean normalized area $\langle
4A/\pi D^2 \rangle$  and (b) the mean normalized perimeter
$\langle P/D \rangle$ of sheetlike plumes.} \label{fig10}
\end{figure}

A striking feature that one can observe from figure \ref{fig3} is
that, as the Rayleigh number is increased, the morphology or
geometry of sheetlike plumes change from more elongated shape to
less elongated and more fragmented. The extracted horizontal cuts
of sheetlike plumes may be characterized by a typical length $l$
and a typical width $w$. With increasing $Ra$, the length of
sheetlike plumes seems to decrease, while the plume's width seems
to increase. To describe these geometric properties
quantitatively, we define the aspect ratio of sheetlike plumes,
$\gamma_{pl}^{sheet}$, as
\begin{equation}
\label{eq:ar} \gamma_{pl}^{sheet} = l/w \mbox{,\ with\ }\left\{ \begin{array}{ll} P = 2(l+w),\\[8pt] A = lw,\\[8pt] l\geq w.
\end{array}\right.
\end{equation}
Figure \ref{fig9} shows the $Ra$-dependency of the mean aspect
ratio of sheetlike plumes $\langle\gamma_{pl}^{sheet}\rangle$ in a
log-log plot. It is seen that $\langle\gamma_{pl}^{sheet}\rangle$
decreases with increasing $Ra$ and the relation can be described
by a power-law with a scaling exponent $-0.23$. We note that
$\langle\gamma_{pl}^{sheet}\rangle$ can also be described by a
logarithmic function of $Ra$ and we cannot definitively conclude
which one, between power-law and logarithm, is a better choice.
Figures \ref{fig10}(a) and (b) show $Ra$-dependence of (a) the
mean normalized area $\langle 4A/\pi D^2 \rangle$ and (b) the mean
normalized perimeter $\langle P/D \rangle$ of sheetlike plumes.
Again, these two quantities are found to decrease with increasing
$Ra$. All these behaviors are well illustrated in figure
\ref{fig3}, and suggest that when the flow becomes more turbulent,
the increased mixing, collision, and convolution can more easily
fragment large-sized sheetlike plumes into smaller ones.

\begin{figure}
\begin{center}
\includegraphics[scale=0.5]{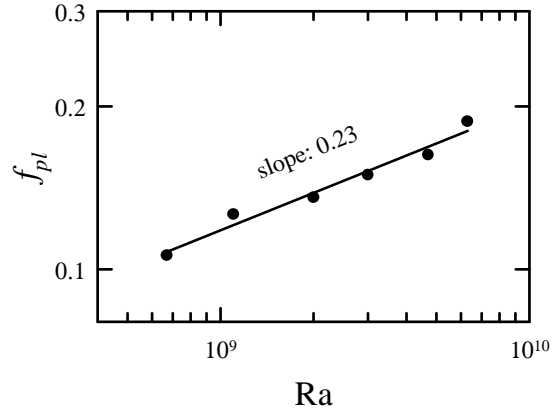}
\end{center}
\caption {The area fraction $f_{pl}$ of sheetlike plumes (area
coverage of plumes over the top plate) as a function of $Ra$ in a
log-log plot. The solid line represents a power-law fit to the
data.} \label{fig11}
\end{figure}

Figure \ref{fig11} shows the plume area fraction $f_{pl}$ as a
function of  the Rayleigh number in a log-log plot. It is seen
that $f_{pl}$ increases with increasing $Ra$ and a power-law
function with a scaling exponent 0.23 can be used to describe well
the relationship between $f_{pl}$ and $Ra$. Using the shadowgraph
technique, \cite{funfschilling2008jfm} studied $Ra$-dependence of
the plume area fraction. Although the scaling range is limited,
they found a power-law scaling of $f_{pl}$ with $Ra$ for
$f_{pl}<0.1$ and the obtained scaling exponent is around 2.

\begin{figure}
\begin{center}
\includegraphics[scale=1]{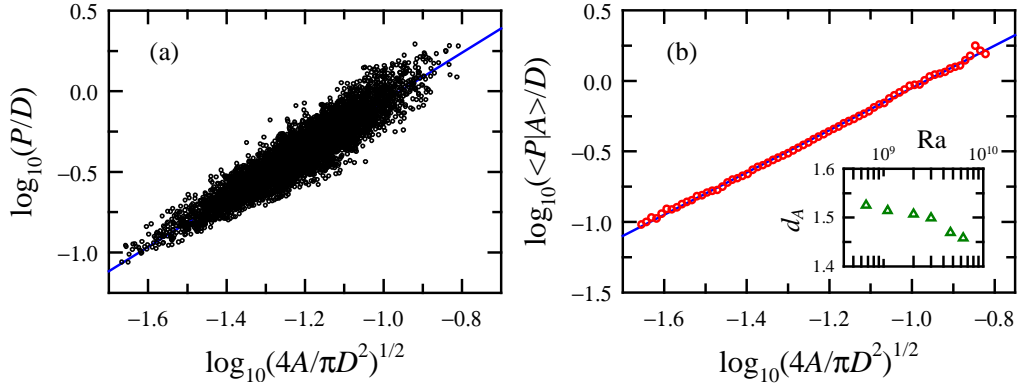}
\end{center}
\caption {(a) Scatter plot of the normalized perimeter $P/D$ vs
the normalized  size $(4A/\pi D^2)^{1/2}$ of sheetlike plumes at
$Ra=3.0\times10^9$. (b) The conditional average $\langle
P|A\rangle/D$ on the normalized size $(4A/\pi D^2)^{1/2}$ for the
same data as (a). Inset is the plot of fractal dimensions of
sheetlike plumes' boundary $d_A$ vs Ra. Solid lines in both (a)
and (b) are power-law fits. } \label{fig12}
\end{figure}

Next we examine the relationship between plume perimeter and area.
Figure \ref{fig12}(a) shows a scatter plot of the normalized
perimeter $P/D$ and the normalized size $(4A/\pi D^2)^{1/2}$ of
sheetlike plumes, which contains a total of $6071$ sheetlike
plumes, extracted from a sequence of $260$ images at Ra
$=3.0\times10^9$. Here, the perimeter and area are normalized by
the diameter and the area of the conducting plate, respectively.
Although the data look somewhat scatter, one sees that the
perimeter increases with area and all data points can be fitted by
a power-low function,
\begin{equation}
\label{eq:plfla} (P/D) \sim (\sqrt{4A/\pi D^2})^{d_A} \mbox{\ \ with\ \ } d_A=1.50.
\end{equation}
To better explore this feature, the conditional average $\langle
P|A\rangle/D$ on the normalized size $(4A/\pi D^2)^{1/2}$ is shown
in figure \ref{fig12}(b). In the figure, a good scaling range can
be seen and the solid line is a power-law fit
$(P/D)=28.1\times(4A/\pi D^2)^{1.50\pm0.02}$. We find that such a
scaling behavior exists for all $Ra$ investigated. The inset of
figure \ref{fig12}(b) shows the $Ra$-dependence of $d_A$. One sees
that these $d_A$ have a mean value of $1.50$, but decrease from
$1.53$ to $1.46$ when $Ra$ increases from $6.7\times10^8$ to
$6.3\times10^9$. These measured values of $d_A$ are further found
to be larger than the value of 1.29 found for isoconcentration
contours of passive scalars measured in the same system
\cite{zhou10passivescalar}.

Mathematically, as introduced by Mandelbrot \cite{mandelbrot1982}
and  Lovejoy \cite{lovejoy1982science}, $d_A$ is the fractal
dimension of the boundary of sheetlike plumes and satisfies $1\leq
d_A < 2$. Fractal dimensions in turbulence have been widely
studied (see, for example, Ref. \cite{sreenivasan1991arfm}). In
the present case, our results seem to suggest that the boundary of
sheetlike plumes is fractal with the dimension of around $1.50$
and the slight decrease of $d_A$ may suggest that the shape of
sheetlike plumes becomes smoother when the flow becomes more
turbulent, as a result of the increased mixing. However, the
extracted sheetlike plumes, as shown in figure \ref{fig3}, do not
look like fractal objects. To understand this and to see if
applying the machinery of fractal analysis to the geometric
properties of plume can shed some light on the problem, we study
the shape complexity of the plumes.

\subsection{Shape complexity of sheetlike plumes}

\begin{figure}
\begin{center}
\includegraphics[scale=1]{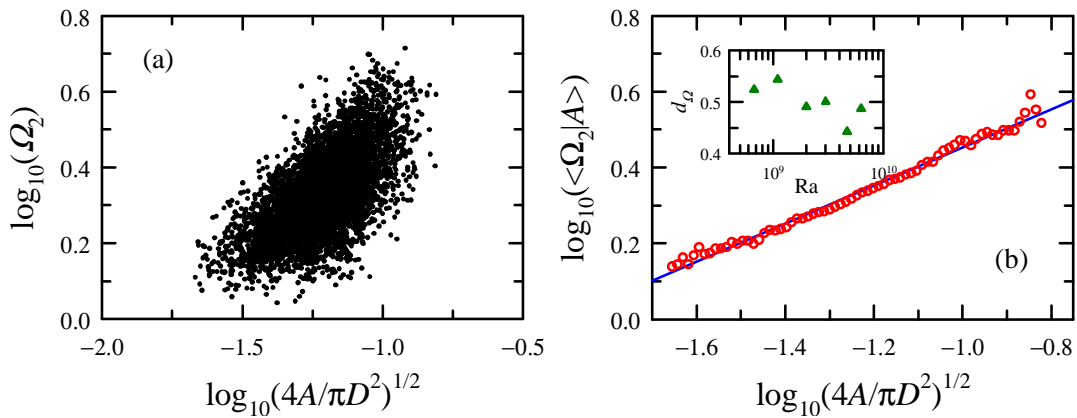}
\end{center}
\caption {(a) Scatter plot of the shape complexity $\Omega_2$ vs
the normalized  size $(4A/\pi D^2)^{1/2}$ of sheetlike plumes at
$Ra=3.0\times10^9$. (b) The conditional average $\langle
\Omega_2|A\rangle$ on the normalized size $(4A/\pi D^2)^{1/2}$
with a solid line as a power-law fit. Inset: $d_{\Omega}$ vs $Ra$.
(same data as in figure \ref{fig9})} \label{fig13}
\end{figure}

The geometric complexity of an object can be characterized by its
shape complexity, which is a dimensionless ratio between its area
and volume. The shape complexity was previously used to study the
shape of isocontours for passive scalar fields in turbulence
\cite{dimotakis1998prl}. We introduce it here to describe the
shape of sheetlike plumes. For a two-dimensional (2D) closed
contour, the shape complexity is defined as follows,
\begin{equation}
\label{eq:sc} \Omega_2=\frac{P}{2\sqrt{\pi A}},
\end{equation}
where $P$ and $A$ are the perimeter of the contour and area
inclosed by the contour,  respectively, and the subscript 2 refers
to 2D. As a circle has the minimum perimeter $2\sqrt{\pi A}$ among
all 2D objects with the same area $A$, the shape complexity
$\Omega_2$ satisfies $1\leq\Omega_2<\infty$ and can be used to
describe the departure of an object from the shape of a circle.
For a fractal object, a larger $\Omega_2$ implies a rougher
contour (or surface) the object. For a non-fractal object, a
larger $\Omega_2$ signifies that the object is more different from
a circle (e.g. it is more elongated).


Figure \ref{fig13}(a) shows a scatter plot of the shape complexity
$\Omega_2$ and the normalized dimension (or size) $(4A/\pi
D^2)^{1/2}$ of sheetlike plumes, obtained from the same data set
as those in figure \ref{fig12}. It is seen that the mean trend of
the relation between the size and shape complexity can be captured
by an increasing function, i.e., the plume with larger size has a
higher probability of possessing a larger $\Omega_2$. To
illustrate this mean trend more clearly, the conditional average
$\langle \Omega_2|A\rangle$ on the normalized size $(4A/\pi
D^2)^{1/2}$ is plotted in figure \ref{fig13}(b). The solid line in
the figure is a power-law fit, i.e.,
\begin{equation}
\label{eq:plfqa} \Omega_2 \sim (\sqrt{4A/\pi D^2})^{d_{\Omega}} \mbox{\ \ with\ \ } d_{\Omega}=0.50\pm0.01
\end{equation}
and the reasonable scaling range can be found. Here, $d_\Omega
\simeq d_A - 1$ is an  immediate consequence of the definition of
shape complexity [Eq. (\ref{eq:sc})] together with the power-law
relation [Eq. (\ref{eq:plfla})]. The inset of figure \ref{fig13}(b) shows the
$Ra$-dependence of $d_{\Omega}$. One sees that $d_{\Omega}$
decreases slightly with increasing $Ra$, which may be understood
from the decrease of $d_A$.

\begin{figure}
\begin{center}
\includegraphics[scale=1]{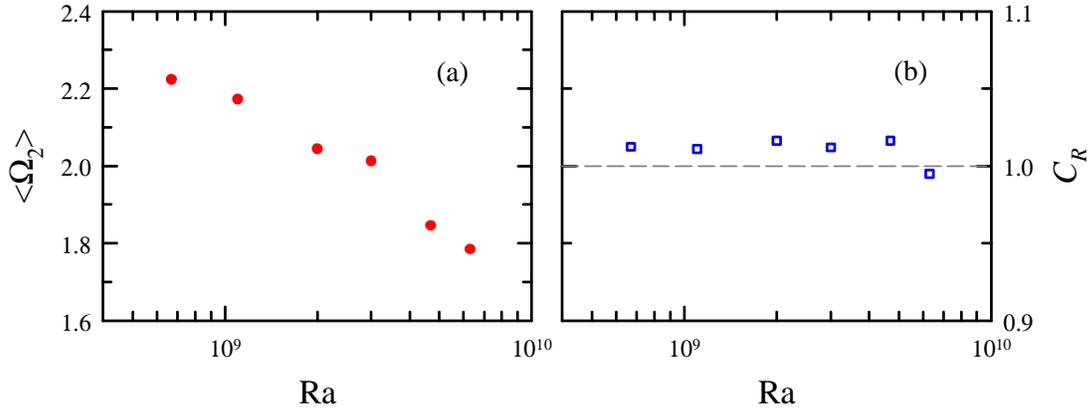}
\end{center}
\caption {(a) $Ra$-dependence of the mean shape complexity
$\langle\Omega_2\rangle$ of  sheetlike plumes. (b) The ratio
between $\langle\Omega_2\rangle$ and
$\frac{1}{\sqrt{\pi}}(\sqrt{\langle\gamma_{pl}^{sheet}\rangle} +
\frac{1}{\sqrt{\langle\gamma_{pl}^{sheet}\rangle}})$, $C_R$, vs $Ra$.}
\label{fig14}
\end{figure}

Figure \ref{fig14}(a) shows the $Ra$-dependence of the mean shape
complexity  $\langle \Omega_2\rangle$. It is seen that $\langle
\Omega_2\rangle$ decreases with increasing $Ra$. This suggests
that, because of the increased mixing, collision and convolution
of sheetlike plumes, the shape of sheetlike plumes become closer
to the shape of a circle when the flow becomes more turbulent.
This also implies the decreases of plume's aspect ratio with
increasing $Ra$. Indeed, with the definitions of
$\gamma_{pl}^{sheet}$ [Eq. (\ref{eq:ar})] and $\Omega_2$ [Eq.
(\ref{eq:plfqa})], one can obtain the relation between of
$\Omega_2$ and $\gamma_{pl}^{sheet}$, i.e.,
\begin{equation}
\label{eq:scar} \Omega_2 = \frac{1}{\sqrt{\pi}}(\sqrt{\gamma_{pl}^{sheet}} + \frac{1}{\sqrt{\gamma_{pl}^{sheet}}}),
\end{equation}
which implies that $\Omega_2$ decreases with decreasing
$\gamma_{pl}^{sheet}$ when  $\gamma_{pl}^{sheet}\geq1$.
Furthermore, $\Omega_2$ is proportional to $\gamma_{pl}^{sheet}$
when $\gamma_{pl}^{sheet}\gg1$. Accordingly, the decrease of
$\langle\Omega_2\rangle$, shown in figure \ref{fig14}(a), may be
understood as the result of the change in aspect ratio of the
sheetlike plumes, i.e. the decrease of
$\langle\gamma_{pl}^{sheet}\rangle$ (see figure \ref{fig9}), as
$\gamma_{pl}^{sheet}$ is much larger than 1 for most of
situations. To see this more clearly, we study the ratio $C_R$
between $\langle\Omega_2\rangle$ and
$\frac{1}{\sqrt{\pi}}(\sqrt{\langle\gamma_{pl}^{sheet}\rangle} +
\frac{1}{\sqrt{\langle\gamma_{pl}^{sheet}\rangle}})$. Note that Eq. (\ref{eq:scar}) is valid only for an individual plume, while $C_R$ is in an average sense. Figure \ref{fig14}(b)
shows $C_R$ as a function of $Ra$. It is seen that $C_R\simeq1$
for nearly all $Ra$ investigated. This result demonstrates clearly
that the decrease of $\Omega_2$ is caused by the decrease of the
plume's aspect ratio, rather than by a decreased roughness in the
case of a fractal object which the sheetlike plumes are not.



\subsection{Distributions of geometric measures of sheetlike plumes}

\begin{figure}
\begin{center}
\includegraphics[scale=1]{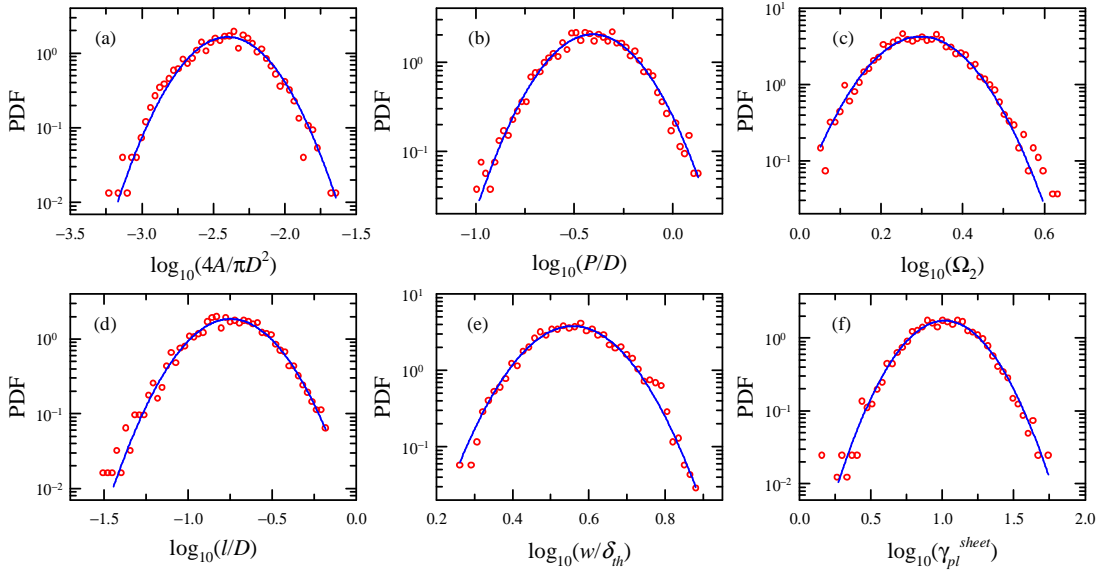}
\end{center}
\caption {PDFs of (a) normalized area $4A/\pi D^2$, (b) normalized
perimeter $P/D$, (c) shape complexity $\Omega_2$, (d) normalized
length $l/D$, (e) normalized width $w/\delta_{th}$, and (f) aspect
ratio $\gamma_{pl}^{rod}$ sheetlike plumes, all obtained at
$Ra=3.0\times10^9$. Solid curves are fittings of log-normal
function to the respective data.} \label{fig15}
\end{figure}

In a previous study, Zhou \emph{et al.} \cite{zhou2007prl} found
that the area of sheetlike plumes obeys log-normal statistics.
Here, we further investigate distributions of other geometric
measures of sheetlike plumes. Figures \ref{fig15}(a)-(f) show the
measured PDFs of normalized geometric measures, i.e. $4A/\pi D^2$,
$P/D$, $\Omega_2$, $l/D$, $w/\delta_{th}$, and
$\gamma_{pl}^{sheet}$, of sheetlike plumes. It is seen that all
these quantities have a same distribution, i.e. the log-normal
distribution. The same distributions were also found for all of
these quantities at other $Ra$ investigated in our experiments.
Together with the log-normal distributions found for mushroomlike
plumes \cite{zhou2002prl}, these findings suggest that the
log-normal distribution is universal for thermal plumes and
log-normal statistics may be used to model them, at least in
turbulent RBC. In addition, this log-normal statistics of thermal
plumes is different from that of passive scalars measured in the
same system, which is found to obey a log-Poisson statistics
\cite{zhou10passivescalar}.


\subsection{Three-dimensional structures of sheetlike plumes}
\label{sec:3D}

\begin{figure}[b]
  \center
\resizebox{0.8\columnwidth}{!}{%
  \includegraphics{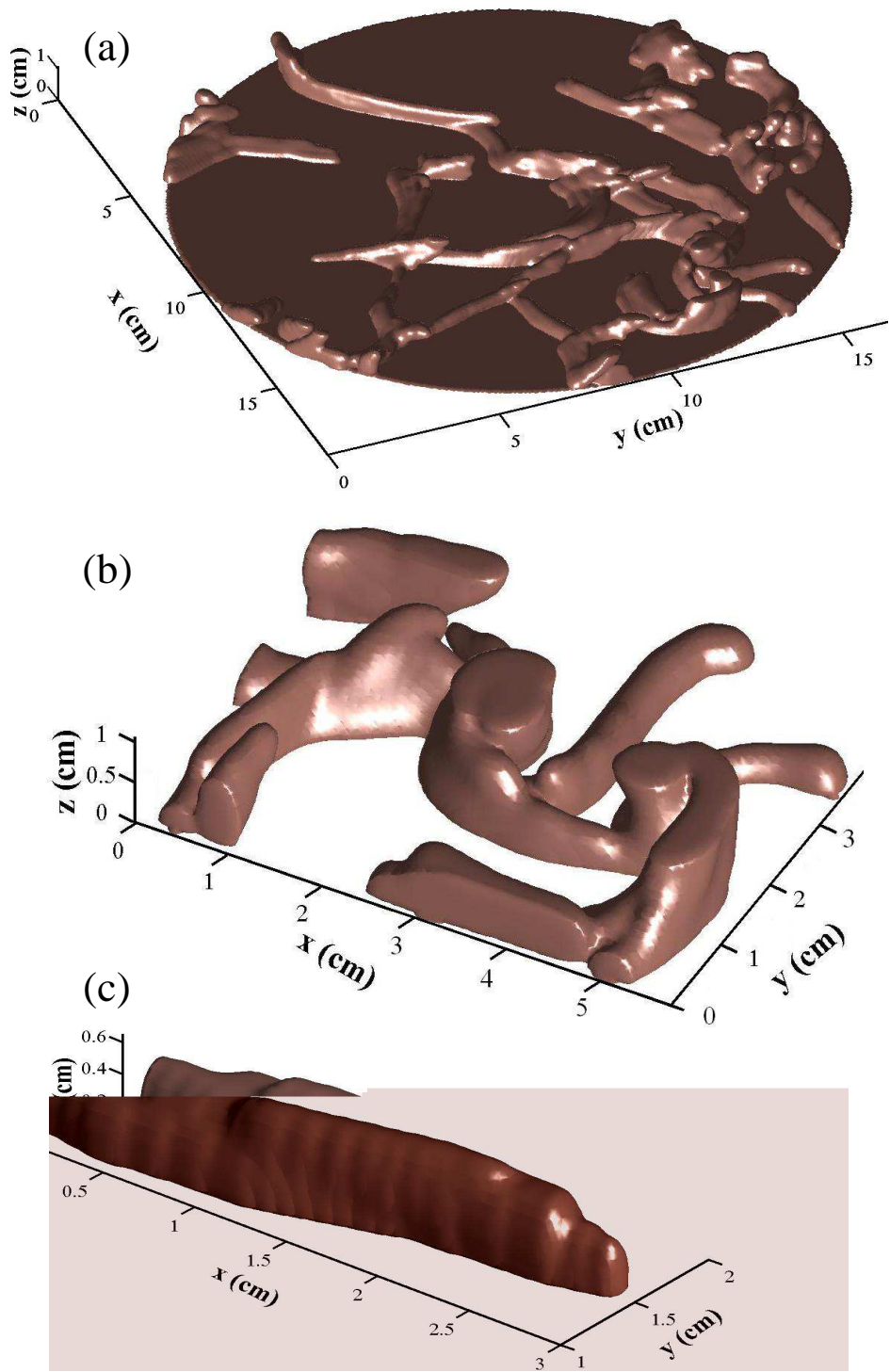}
}\caption{(a) Three-dimensional sheetlike/rodlike thermal plumes
obtained near the top plate by tomographic reconstruction
technique ($Ra=2.0\times10^9$, $\Gamma=1$). (b) An enlarged region
of (a), showing the morphological transformation of rodlike plumes
into mushroomlike ones through convolution/spiraling. (c) An
individual rodlike plume.}
 \label{fig16}
\end{figure}

Finally, we study the 3D structures of sheetlike plumes. Results
obtained in  section \ref{sec:z} have suggested that the vertical
extent of what is called sheetlike plumes is only of the order of
the thermal boundary layer thickness, and hence not large enough
to form a sheet. To see this more clearly , we used a tomographic
reconstruction technique to construct the 3D image of thermal
plumes from sequences of 2D images acquired near the top plate of
the cell. To achieve this, the convection cell was placed on a
translational stage and, as the cell traverses continuously at a
speed of $\sim 1 cm/sec$, a series of photographs of TLC
microspheres were recorded by a Nikon D3 camera (3CCD, with a
resolution of $2397\times1591$ pixels) operating at 11 frames/s.
As the speed of the the horizontal motion of the sheetlike plumes
are about $0.4 cm/sec$, the sequences of the horizontal slices of
the plumes may be regarded as taken at approximately the same
time. In post-experiment analysis,  2D horizontal cuts of cold
plumes were extracted from the images taken in each run and a
MATLAB script was used to reconstruct 3D thermal plumes from these
extracted 2D cuts. Figure \ref{fig16}(a) shows an example of the
reconstruction of 3D thermal plumes near the top plate. From the
figure, one can see how mushroom-like plumes are formed by the
convolution (or spiraling) of sheetlike plumes. Figure
\ref{fig16}(b) shows this process more clearly, which is an
enlarged region of the image. Another noteworthy feature is that
most plumes have a one-dimensional structure, rather than a
sheetlike shape. This geometric feature may be illustrated more
clearly by figure \ref{fig16}(c), which shows the reconstruction
of an enlarged individual 3D plume. It is seen that both the
height and the width of this plume extend only to a few
millimeters, while its length extends to a few centimeters. This
is the most direct evidence thus far that shows thermal plumes
near the conducting plates are only one-dimensional structures
with horizontal length being much larger than their horizontal
width and vertical extent, which has been suggested previously
\cite{funfschilling2008jfm}. Therefore, we may hereafter call
thermal structures near the conducting plates rod-like plumes.
Future investigations will be focused on the geometric properties
of 3D structures of both rod-like and mushroom-like thermal
plumes.

\section{Conclusions}

In this paper, we have presented a detailed experimental study of
the temperature and velocity fields in turbulent
Rayleigh-B\'{e}nard convection using the
thermochromic-liquid-crystal technique. The number statistics and
geometric properties of sheetlike/rodlike thermal plumes were
investigated. Major findings are summarized as follows:

(i) When observed from above, the previous-called sheetlike
thermal plumes near the top plate seem to be only one-dimensional
structures and should be called rodlike plumes hereafter. These
plumes evolve morphologically, i.e. convolute or spiral, to
mushroomlike plumes. The width ($\ell$) of the region near the
plates within which these rodlike plumes exist and evolve
morphologically is only associated with the thermal boundary
thickness (i.e. $\ell \sim$ several $\delta_{th}$).

(ii) The numbers of sheetlike/rodlike and mushroomlike plumes,
$N_{pl}^{sheet}$ and $N_{pl}^{mush}$, are found to both scale as
$Ra^{0.3}$. This finding suggests that the total amount of heat
flow is dominated only by the number of thermal (rodlike or
mushroomlike) plumes and the normalized mean heat content carried
by each type of plumes is approximately independent on $Ra$.

(iii) As the turbulent intensity is increased, large-sized sheetlike/rodlike plumes are more easily to be fragmented into smaller ones and hence their aspect ratios decrease.

(iv) Although sheetlike/rodlike plumes do not look like fractal objects, , power-law relations can be
used to characterize the relationship between their perimeter and
area and between their size and shape complexity.

(v) Geometric measures of thermal plumes are found to have a
universal distribution,  i.e. the log-normal distribution. Thus,
log-normal statistics may be used to model thermal plumes. In
contrast, for the mixing of passive scalars measured in the same
system, it is found that the same quantities obey log-Poisson
statistics \cite{zhou10passivescalar}.

(vi) The Ra-dependency of the plume area density ($f_r \sim
Ra^{0.23}$)suggests that the plume number increase with $Ra$ ($N
\sim Ra^{0.3}$) can be largely attributed to the increased
emission of plumes (as $Ra$ is increased), rather than as a result
of fragmentation of plumes.

\section*{Acknowledgments}

We gratefully acknowledge support of this work by the Research Grants Council of Hong
Kong SAR (Nos. CUHK403806 and 403807) (K.Q.X.) and by the Natural Science Foundation of Shanghai (No. 09ZR1411200), ``Chen Guang" project (No. 09CG41), and RFDP of Ministry of Education of China (No. 20093108120007) (Q.Z.).


\section*{References}
\begin{thebibliography}{}

\bibitem{siggia1994arfm} Siggia E D 1994 High Rayleigh number convection {\it Annu. Rev. Fluid Mech.} {\bf 26} 137-168

\bibitem{agl} Ahlers G, Grossmann S and Lohse D 2002 Heat transfer and large scale dynamics in turbulent Rayleigh-B\'{e}nard convection {\it Rev. Mod. Phys.} {\bf 81} 503-537

\bibitem{lx} Lohse D and Xia K Q 2010 Small scale properties of turbulent Rayleigh-B\'{e}nard convection {\it Annu. Rev. Fluid Mech.} {\bf 42} 335-364

\bibitem{morgan1968jgr} Morgan W J 1968 Rises trenches great faults and crustal blocks {\it J. Geophys. Res.} {\bf 73} 1959

\bibitem{jellinek2004rog} Jellinek A M and Manga M 2004 Links between long-lived hot spots, mantle plumes, D'', and plate tectonics {\it Rev. Geophys.} {\bf 42} RG3002

\bibitem{rieutord1995aaa} Rieutord M and Zahn J P 1995 Turbulent plumes in stellar convective envelopes {\it Astron. Astrophys.} {\bf 296} 127-138

\bibitem{shang2003prl} Shang X D, Qiu X L, Tong P and Xia K Q 2003 Measured local heat transport in turbulent Rayleigh-B\'{e}nard convection {\it Phys. Rev. Lett.} {\bf 90} 074501

\bibitem{shang2004pre} Shang X D, Qiu X L, Tong P and Xia K Q 2004 Measurements of the local convective heat flux in turbulent Rayleigh-B\'{e}nard convection {\it Phys. Rev. E} {\bf 70} 026308

\bibitem{pinton2007prl} Gasteuil Y, Shew W L, Gibert M, Chilla F, Castaing B and Pinton J F 2007 Lagrangian temperature, velocity, and local heat flux measurement in Rayleigh-B\'{e}nard convection {\it Phys. Rev. Lett.} {\bf 99} 234302

\bibitem{gl2004pof} Grossmann S and Lohse D 2004 Fluctuations in turbulent Rayleigh-B\'{e}nard convection: The role of plumes {\it Phys. Fluid} {\bf 16} 4462-4472

\bibitem{moses1993jfm} Moses E, Zocchi G and Libchaber A 1993 An experimental study of laminar plumes {\it J. Fluid Mech.} {\bf 251} 581-601

\bibitem{kaminski2003jfm} Kaminski E and Jaupart C 2003 Laminar starting plumes in high-Prandtl-number fluids {\it J. Fluid Mech.} {\bf 478} 287-298

\bibitem{zocchi1990pa} Zocchi G, Moses E and Libchaber A 1990 Coherent structures in turbulent convection: an experimental study {\it Physica (Amsterdam)} {\bf A166} 387¨C407

\bibitem{moses1991epl} Moses E, Zocchi G, Procaccia I and Libchaber A 1991 The Dynamics and Interaction of Laminar Thermal Plumes {\it Europhys. Lett.} {\bf 14} 55-60

\bibitem{gluckman1993pof} Gluckman B J, Willaime H and Gollub J P 1993 Geometry of isothermal and isoconcentration on surfaces in thermal turbulence {\it Phys. Fluids A} {\bf 5} 647-661

\bibitem{ciliberto1996pre} Ciliberto S, Cioni S and Laroche C 1996 Large-scale flow properties of turbulent thermal convection {\it Phys. Rev. E} {\bf 54} R5901-5904

\bibitem{zhang1997pof} Zhang J, Childress S and Libchaber A 1997 Non-Boussinesq effect: Thermal convection with broken symmetry {\it Phys. Fluids} {\bf 9} 1034-1042

\bibitem{du1998prl} Du Y B and Tong P 1998 Enhanced heat transport in turbulent convection over a rough surface {\it Phys. Rev. Lett.} {\bf 81} 987-990

\bibitem{qiu2001prl} Qiu X L and Tong P 2001 Onset of convection in turbulent Rayleigh-B\'{e}nard convection {\it Phys. Rev. Lett.} {\bf 87} 094501

\bibitem{xi2004jfm} Xi H D, Lam S and Xia K Q 2004 From laminar plumes to organized flows: the onset of large-scale circulation in turbulent thermal convection {\it J. Fluid Mech.} {\bf 503} 47-56

\bibitem{zhou2002prl} Zhou S Q and Xia K Q 2002 Plume statistics in thermal turbulence: mixing of an active Scalar {\it Phys. Rev. Lett.} {\bf 89} 184502

\bibitem{zhou2009} Zhou S Q and Xia K Q 2009 Statistical characterization of the thermal structures in turbulent Rayleigh-B\'{e}nard {\it to be published}

\bibitem{tanaka1980ijhmt} Tanaka H and Miyata H 1980 Turbulent natural convection in a horizontal water layer heated from below {\it Int. J. Heat Mass Transfer} {\bf 23} 1273-1281

\bibitem{vorobie2002jfm} Vorobie P and Ecke R 2002 Turbulent rotating convection: an experimental study {\it J. Fluid Mech.} {\bf 458} 191-218

\bibitem{funfschilling2004prl} Funfschilling D and Ahlers G 2004 Plume motion and large-scale circulation in a cylindrical Rayleigh-B\'{e}nard cell {\it Phys. Rev. Lett.} {\bf 92} 194502

\bibitem{haramina2004pre} Haramina T and Tilgner A 2004 Coherent structures in boundary layers of Rayleigh-B\'{e}nard convection {\it Phys. Rev. E} {\bf 69} 056306

\bibitem{puthenveettil2005jfm1} Puthenveettil B A and Arakeri J H 2005 Plume structure in high-Rayleigh-number convection {\it J. Fluid Mech.} {\bf 542} 217-249

\bibitem{puthenveettil2005jfm2} Puthenveettil B A and Arakeri J H 2005 The multifractal nature of plume structure in high-Rayleigh-number convection {\it J. Fluid Mech.} {\bf 526} 245-256

\bibitem{zhou2007prl} Zhou Q, Sun C and Xia K Q 2007 Morphological evolution of thermal plumes in turbulent Rayleigh-B\'{e}nard convection {\it Phys. Rev. Lett.} {\bf 98} 074501

\bibitem{funfschilling2008jfm} Funfschilling D, Brown E and Ahlers G 2008 Torsional oscillations of the large-scale circulation in turbulent Rayleigh-B\'{e}enard convection {\it J. Fluid Mech.} {\bf 607} 119-139

\bibitem{puthenveettil2008jfm} Puthenveettil B A and Arakeri J H 2008 Convection due to an unstable density difference across a permeable membrane {\it J. Fluid Mech.} {\bf 609} 139-170

\bibitem{shishkina2008jfm} Shishkina O and Wagner C 2008 Analysis of sheet-like plumes in turbulent Rayleigh-B\`{e}nard convection {\it J. Fluid Mech.} {\bf 599} 383-404

\bibitem{xi2006pre} Xi H D, Zhou Q and Xia K Q 2006 Azimuthal motion of the mean wind in turbulent thermal convection {\it Phys. Rev. E} {\bf 73} 056312

\bibitem{lui1997} Lui S L 1997 Experimental investigation of the temperature field in turbulent convection {\it M.Phil. Thesis} The Chinese University of Hong Kong

\bibitem{lui1998pre} Lui S L and Xia K Q 1998 Spatial structure of the thermal boundary layer in turbulent convection {\it Phys. Rev. E} {\bf 57} 5494-5503

\bibitem{castaing1989jfm} Castaing B, Gunaratne G, Heslot F, Kadanoff L, Libchaber A, Thomae S, Wu X Z, Zaleski S and Zanetti G 1989 Scaling of hard thermal turbulence in Rayleigh-B\'{e}nard convection {\it J. Fluid Mech.} {\bf 204} 1-30

\bibitem{xia2002prl} Xia K Q, Lam S and Zhou S Q 2002 Heat-flux measurement in high-Prandtl-number turbulent Rayleigh-B\'{e}nard convection {\it Phys. Rev. Lett.} {\bf 88} 064501

\bibitem{ahlers2003prl} Nikolaenko A and Ahlers G 2003 Nusselt number measurements for turbulent Rayleigh-B\'{e}nard convection {\it Phys. Rev. Lett.} {\bf 91} 084501

\bibitem{dimotakis1998prl} Catrakis H J and Dimotakis P E 1998 Shape complexity in turbulence {\it Phys. Rev. Lett.} {\bf 80} 968-971

\bibitem{zhou10passivescalar} Zhou Q and Xia K Q 2010 Mixing evolution and geometric properties of passive scalar field in high-Schmidt-number buoyancy-driven turbulence {\bf to be published}

\bibitem{mandelbrot1982} {\it Mandelbrot B B 1982 The Fractal Geometry of Nature} (Freeman, NY, 1982)

\bibitem{lovejoy1982science} Lovejoy S 1982 Area-perimeter relation for rain and cloud areas {\it Science} {\bf 216} 185-187

\bibitem{sreenivasan1991arfm} Sreenivasan K R 1991 Fractals and multifractals in fluid turbulence {\it Annu. Rev. Fluid Mech.} {\bf 23} 539-600.


%
%
%

\end{thebibliography}
\end{document}